\documentclass[useAMS,usenatbib]{mn2e}

\usepackage{epsfig}
\usepackage{amsmath}
\usepackage{amssymb}
\usepackage{natbib}
\usepackage{threeparttable} 
\usepackage{booktabs}

\usepackage{epstopdf}
\usepackage{times}

\newcommand{\chan}{\textit{Chandra}}
\newcommand{\swift}{\textit{Swift}}
\newcommand{\rxte}{\textit{RXTE}}

\newcommand{\maxi}{\textit{MAXI}}

\newcommand{\Msun}{\mathrm{M}_{\odot}}
\newcommand{\lum}{\mathrm{erg~s}^{-1}}
\newcommand{\flux}{\mathrm{erg~cm}^{-2}~\mathrm{s}^{-1}}

\newcommand{\cnts}{\mathrm{c~s}^{-1}}
\newcommand{\mdot}{\mathrm{M_{\odot}~yr}^{-1}}
\newcommand{\mdotgs}{\mathrm{g~s}^{-1}}
\newcommand{\nh}{\mathrm{cm}^{-2}}

\newcommand{\dist}{(D/4.7~\mathrm{kpc})^2}

\newcommand{\dens}{\mathrm{g~cm}^{-3}}

\newcommand{\source}{HETE~J1900.1--2455}
\newcommand{\exo}{EXO 0748--676}

\newcommand{\ks}{KS~1731--260}

\def \mnras {MNRAS}
\def \apj {ApJ}
\def \apjs {ApJS}
\def \apjl {ApJ}
\def \aap {A\&A}
\def \nat {Nature}

\def \atel {Astron. Telegram}

\def \pasj {PASJ}

\def \prc {Phys. Rev. C}

\def \pre {Phys. Rev. E}
\def \prl {Phys. Rev. Lett.}

\def \apss {A\&AS}

\title[\source\ in quiescence]{A cold neutron star in the transient low-mass X-ray binary \source\ after 10 years of active accretion}
\author[Degenaar et al.]
{
N. Degenaar$^{1,2}$\thanks{e-mail: degenaar@ast.cam.ac.uk}, L.S. Ootes$^{2}$, M.T. Reynolds$^{3}$, R. Wijnands$^{2}$, and D. Page$^{4}$\\
$^1$Institute of Astronomy, University of Cambridge, Madingley Road, Cambridge CB3 OHA, UK\\
$^2$Anton Pannekoek Institute for Astronomy, University of Amsterdam, Science Park 904, 1098 XH, Amsterdam, the Netherlands\\
$^3$Department of Astronomy, University of Michigan, 1085 South University Avenue, Ann Arbor, MI  48109, USA\\
$^4$Instituto de Astronom\'{i}a, Universidad Nacional Aut\'{o}noma de M\'{e}xico, Mexico D.F. 04510, Mexico
}

\begin{document}

\date{Accepted 2016 September 23. Received 2016 September 23; in original form 2016 July 12}

\pagerange{\pageref{firstpage}--\pageref{lastpage}} \pubyear{0000}

\maketitle

\label{firstpage}

\begin{abstract}
The neutron star low-mass X-ray binary and intermittent millisecond X-ray pulsar \source\ returned to quiescence in late 2015, after a prolonged accretion outburst of $\simeq$10~yr. Using a \chan\ observation taken $\simeq$180~d into quiescence we detect the source at a luminosity of $\simeq$$4.5\times10^{31}~\dist~\lum$ (0.5--10 keV). The X-ray spectrum can be described by a neutron star atmosphere model with a temperature of $\simeq$54~eV for an observer at infinity. We perform thermal evolution calculations based on the 2016 quiescent data and a $\lesssim$98~eV temperature upper limit inferred from a \swift\ observation taken during an unusually brief ($\lesssim$2~weeks) quiescent episode in 2007. We find no evidence in the present data that the thermal properties of the crust, such as the heating rate and thermal conductivity, are different than those of non-pulsating neutron stars. Finding this neutron star so cold after its long outburst imposes interesting constraints on the heat capacity of the stellar core; these become even stronger if further cooling were to occur.

\end{abstract}


\section{Introduction}

Neutron stars are one of the possible remnants of once massive stars that ended their life in a supernova explosion. A defining property of neutron stars is that they are very compact; despite having a mass of $\simeq$1.4$~\Msun$, their radius is only $\simeq$10~km. As a result, their interior density rises beyond the density of atomic nuclei. Neutron stars are therefore of prime interest to understand the properties of ultra-dense matter \citep[e.g.][for a review]{lattimer2011}.

Our Galaxy harbors $>$100 neutron stars that are part of low-mass X-ray binaries (LMXBs) and accrete gas from a $\lesssim$1~$\Msun$ companion star via an accretion disk. Many of these systems are transient; during X-ray luminous phases matter is rapidly falling toward the neutron star, but during intervening quiescent phases the accretion rate, and hence the X-ray luminosity, is strongly reduced. Two sub-classes of neutron star LMXBs are the quasi-persistent sources and the accreting millisecond X-ray pulsars (AMXPs). Each make up $\simeq$10 per cent of the current population of neutron star LMXBs, and each have distinct outburst and quiescent properties. 

Quasi-persistent LMXBs stand out by showing prolonged accretion outbursts of years to decades rather than weeks to months. Moreover, several of these sources show strong thermal emission in quiescence that gradually decreases on a timescale of years \citep[e.g.][]{cackett2013_1659,fridriksson2011,degenaar2014_exo3,homan2014,merritt2016}. The crust of a neutron star is heated during accretion outbursts via pycnonuclear fusion reactions that take place at $\simeq$1~km depth, and electron captures occurring at shallower depth. Together, these processes deposit an energy $\simeq1-2$~MeV per accreted nucleon, heating the crust \citep[e.g.][]{haensel1990a,haensel2008,brown1998}. The temperature evolution in quiescence of five quasi-persistent sources can be successfully explained as cooling of the strongly-heated neutron star crust and offers valuable insight into the structure and composition of these neutron stars \citep[e.g.][]{rutledge2002,wijnands2002,wijnands2004,shternin07,brown08,page2013,medin2014,deibel2015,horowitz2015,turlione2013,cumming2016}.\footnote{Recent studies have also revealed crust cooling in three neutron stars with short outbursts \citep[e.g.][]{degenaar2011_terzan5_3,degenaar2015_ter5x3,waterhouse2016}.} 

AMXPs distinguish themselves by displaying coherent X-ray pulsations when accreting \citep[e.g.][]{wijnands1998}. It is believed that in these objects the stellar magnetic field is strong enough to disrupt the accretion flow and channel plasma to the magnetic poles of the rapidly rotating neutron star. In quiescence, the persistently-pulsating AMXPs show weak or no thermal X-rays but strong power-law emission \citep[e.g.][]{campana2005_amxps,campana2008,jonker2005,wijnands05_amxps,heinke2009,degenaar2012_amxp}. Such a hard emission component is also seen in the quiescent spectra of some non-pulsating neutron star LMXBs, though it is typically much less prominent. The hard quiescent X-rays are often ascribed to residual accretion or non-thermal emission processes related to the neutron star magnetic field \citep[e.g.][]{campana1998,rutledge2001,degenaar2012_amxp,chakrabarty2014_cenx4,wijnands2014}.

\source\ shares properties of both the aforementioned sub-classes of neutron star LMXBs; it is the only known AMXP that accreted for a full decade, and the only quasi-persistent source that acted as an AMXP. The source was first seen in outburst in 2005 June when it exhibited a thermonuclear X-ray burst \citep[][]{vanderspek2005}. X-ray pulsations at a frequency of 377.3~Hz were soon found, but were detected only sporadically after $\simeq$2 months \citep[][]{galloway2008_hete,patruno2012_hete}. The source is therefore referred to as an ``intermittent AMXP''. X-ray burst analysis suggests a source distance of $D=4.7$~kpc \citep[][]{galloway06}. 

The $\simeq$10-year long outburst of \source\ ended in late 2015. We report on a \chan\ ToO observation obtained $\simeq$180~d after it went quiescent. We combine the obtained temperature measurement with an upper limit from \swift\ in 2007 when the source disappeared for $\lesssim$2 weeks, to perform thermal evolution simulations and to probe the thermal properties of this neutron star.


\section{Observations and data analysis}

We performed spectral fits in \textsc{XSpec} (v. 12.9). Interstellar absorption was included by using \textsc{tbabs} with \textsc{vern} cross-sections \citep[][]{verner1996} and \textsc{wilm} abundances \citep[][]{wilms2000}. Fluxes were calculated using \textsc{cflux}. Quoted errors reflect 1$\sigma$ confidence intervals and upper limits are given at 95\% confidence level.

\subsection{\chan\ quiescence observation in 2016}

We observed \source\ with \chan\ for $\simeq$29.7~ks from 2016 April 18 at 21:50 UT till April 19 at 06:48 UT. The source was placed on the ACIS S-3 chip, which was operated in very faint, timed mode. We reduced and analyzed the data using \textsc{ciao} v. 4.8. Source events were extracted from a circular region with a radius of $2''$ and a surrounding source-free annulus with an inner--outer radius of $5''$--$25''$ was used for the background. The source was detected at a net count rate of $(2.36\pm 0.03)\times 10^{-3}~\cnts$, yielding a total of 70 source photons. Spectra and response files were extracted using \textsc{specextract}. We grouped the spectral data to $>$1~counts~bin$^{-1}$ using \textsc{grppha} and then fitted the data in the 0.3--7 keV range applying W-statistics \citep[i.e. Cash-statistics with background subtraction;][]{wachter1979}.

\subsection{\swift/XRT non-detections in 2007 and 2016}

Between 2016 March 7 and April 9, \swift\ observed \source\ several times with the XRT in photon counting mode (ObsID 00030946021--27). We use these observations, with exposure times of $\simeq$0.9--1.1~ks, to obtain constraints on the neutron star temperature prior to our \chan\ observation. We also re-analyze a single 0.8~ks exposure obtained on 2007 June 5 \citep[obsID 00030946002;][]{deeg07_hetenon}. None of these \swift\ observations detected the source. The XRT data were analyzed using standard tools incorporated in \textsc{heasoft} (v. 6.18). The data were first reduced using the reprocessing pipeline and then examined using \textsc{XSelect}. Ancillary response files were created using \textsc{xrtmkarf}, and the response matrix file (v. 15) was obtained from the calibration data base. 


\begin{figure}
\includegraphics[width=8.5cm]{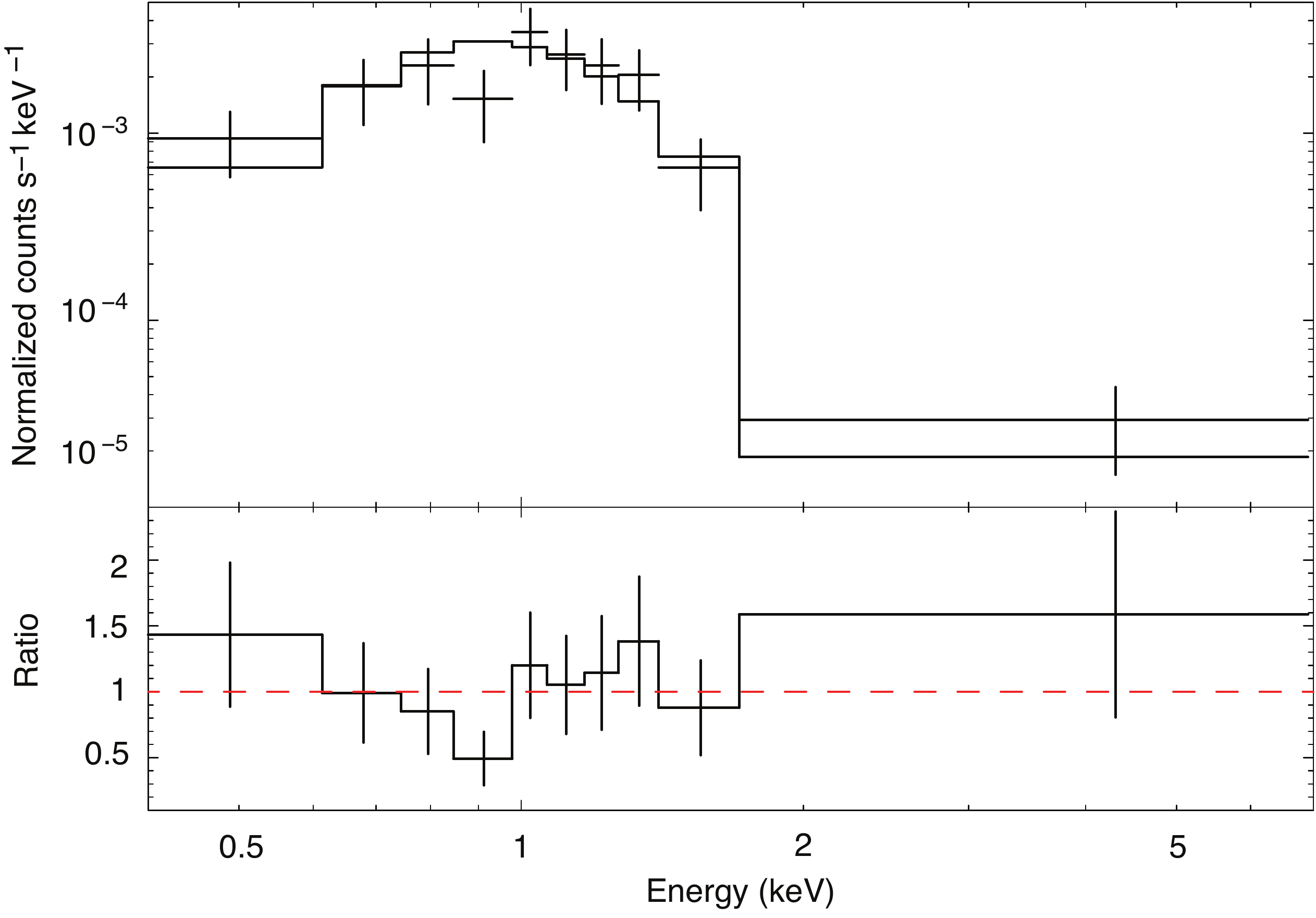}
\caption[]{\chan\ spectral data and fit to \textsc{tbabs*nsatmos} (rebinned for visual clarity). The lower panel shows the data to model ratio. 
}
 \label{fig:spec}
\end{figure}   

\section{Observational results}

\subsection{\chan: A thermal spectrum and a cold neutron star}\label{subsec:chan}

When fitting the \chan\ spectral data with a simple power-law model (\textsc{pegpwrlw}), we obtain an extremely steep index of $\Gamma=5.2\pm0.9$. This suggests that the spectrum has a soft, thermal shape. We therefore modeled the spectral data with a physically-motivated neutron star atmosphere model, for which we chose \textsc{nsatmos} \citep[][]{heinke2006}. In this model we fixed $M=1.4~\Msun$, $R=10$~km, $D=4.7$~kpc and assumed that the entire surface was emitting (i.e. the normalization was set to 1). 
We note that due to the low flux, we cannot determine the presence of hotspots, which would result in an overestimate of the neutron star surface temperature \citep[e.g.][]{elshamouty2016}. 

The \textsc{nsatmos} model can adequately describe the spectral data, as is shown in Figure~\ref{fig:spec}. We find $N_{\mathrm{H}} =  (2.2 \pm 0.7) \times 10^{21}~\nh$ and measure a neutron star temperature, as seen by an observer at infinity, of $kT^{\mathrm{\infty}}_{\mathrm{eff}}=54.4\pm1.7$~eV. The inferred 0.5--10 keV flux of $F_{\mathrm{X}} = (1.7 \pm 0.8) \times 10^{-14}~\flux$ translates into a luminosity of $L_{\mathrm{X}}=(4.5 \pm 2.1) \times10^{31}~\lum$ at 4.7~kpc. We estimate a thermal bolometric flux of $F^{\mathrm{q}}_{\mathrm{bol}} = (2.3 \pm 1.4) \times 10^{-14}~\flux$ by extrapolating the fit to the 0.01--100~keV range, which yields $L^{\mathrm{q}}_{\mathrm{bol}} = (6.1 \pm 3.7) \times 10^{31}~\lum$.

To set an upper limit on the contribution from a hard emission tail, we added a power-law component (\textsc{pegpwrlw}) to our spectral model. Given the few counts, we fixed the power-law index to $\Gamma=1$ and~2. This suggests that any hard spectral component contributes $\lesssim$9 per cent to the total unabsorbed 0.5--10 keV flux. For these fits the neutron star temperature decreases by $\simeq$1~eV.

\subsection{\swift: Temperature upper limits in quiescence}\label{subsec:swiftresults}

For each \swift/XRT non-detection we counted the number of photons in a circular $10''$ region centered on the source position and calculated a 95\% confidence upper limit by using the tables of \citet{gehrels1986}. For each data set we simulated \textsc{nsatmos} spectra with $D=4.7$~kpc, $M=1.4~\Msun$, $R=10$~km, $N_{\mathrm{H}}=2\times10^{21}~\nh$, and different temperatures, using the observation-specific response files. A temperature upper limit was then estimated by matching the determined XRT count rate upper limit with that predicted by the simulated spectra. 

We obtain $kT^{\mathrm{\infty}}_{\mathrm{eff}} \lesssim 72-93$~eV for the different \swift\ observations of 2016. The corresponding upper limits on the 0.5--10 keV unabsorbed fluxes and luminosities are $F_{\mathrm{X}} \lesssim (1-4)\times 10^{-13}~\flux$ and $L_{\mathrm{X}}\lesssim (2.6-10.6)\times 10^{32}~\dist~\lum$, respectively. Summing all XRT data (6.5~ks) results in a non-detection that suggests $kT^{\mathrm{\infty}}_{\mathrm{eff}} \lesssim63$~eV and $F_{\mathrm{X}}\lesssim 6\times 10^{-14}~\flux$.

\begin{figure}
\includegraphics[width=8.5cm]{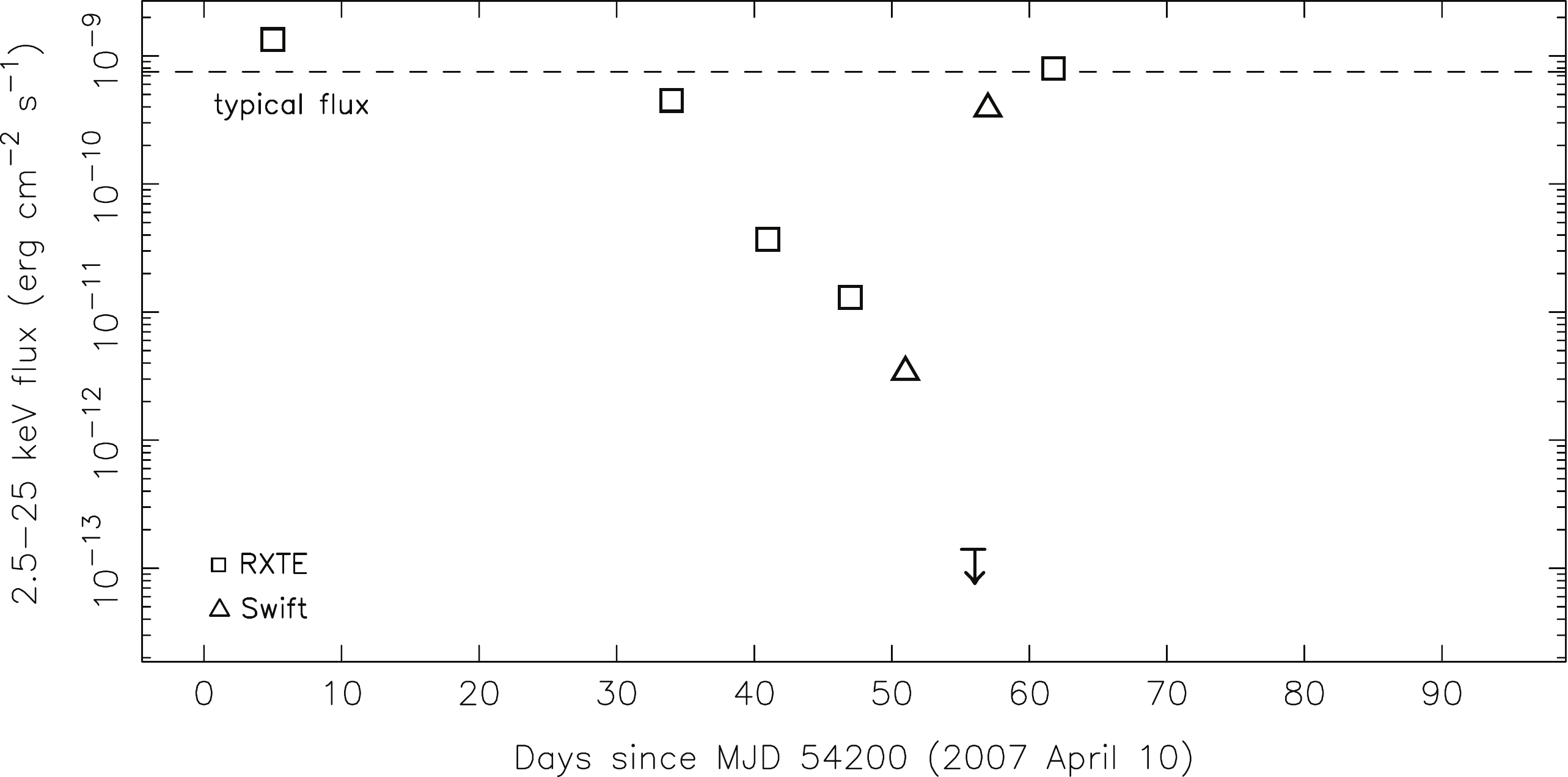}\vspace{+0.2cm}
\includegraphics[width=8.5cm]{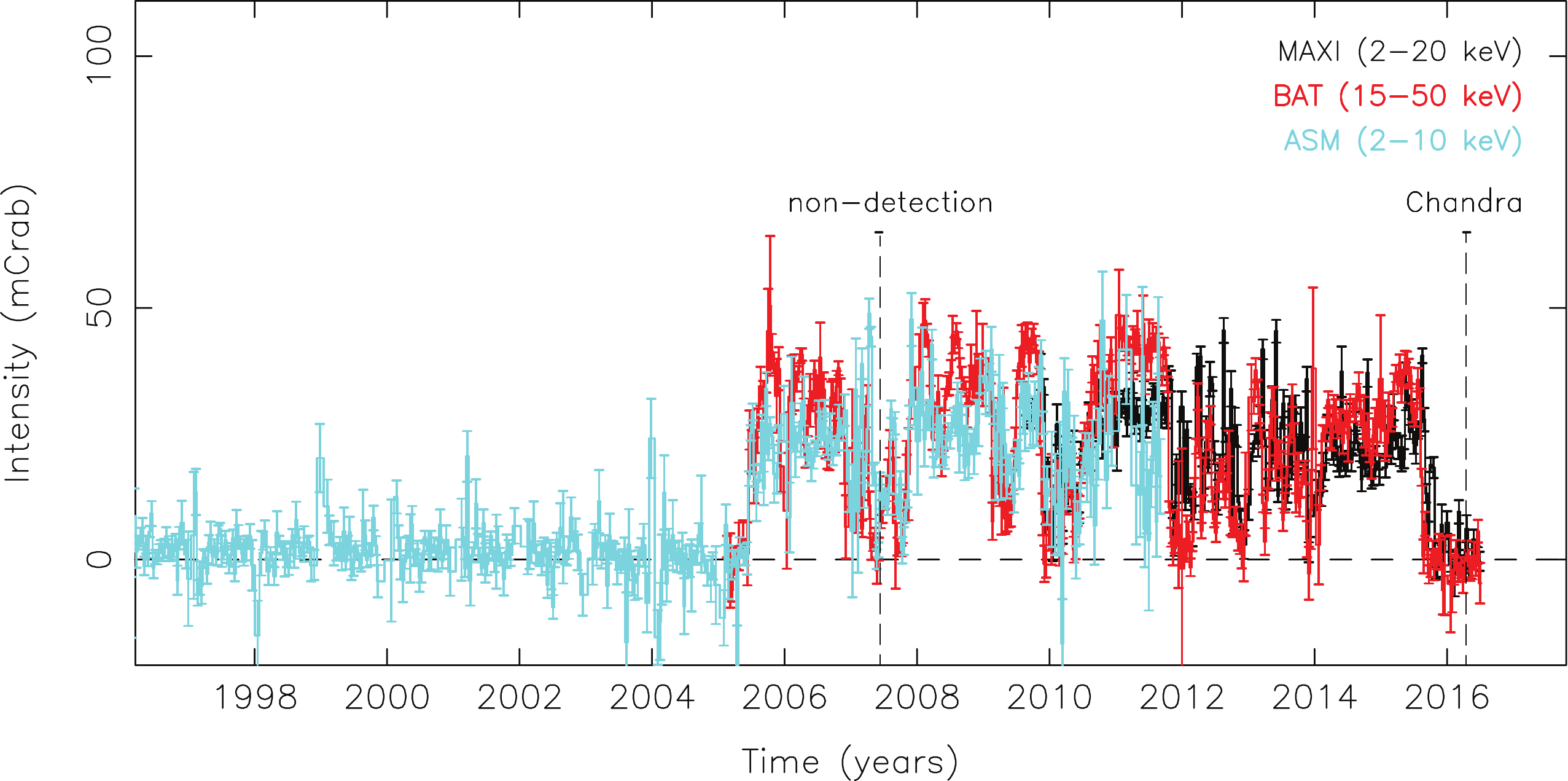}
\caption[]{Light curves of \source. Top: Combined \rxte/PCA and \swift/XRT light curve in 2007 April--June \citep[fluxes from][]{galloway2007_hete,deeg07_hetedecay,degenaar2007_heterecover,deeg07_hetenon}. Bottom: \rxte/ASM, \swift/BAT, and \maxi\ monitoring data binned per 10 days. 
}
 \label{fig:combined_lc}
 \end{figure}

In 2007 May, the flux of \source\ decayed from its regular outburst level by $>$2 orders of magnitudes in $\simeq$22~d \citep[][]{galloway2007_hete,deeg07_hetedecay}, leading to a non-detection with \swift/XRT on June 5 \citep[][]{deeg07_hetenon}. Only $\simeq$5~d later, however, the source was found back at its outburst level \citep[][]{degenaar2007_heterecover}. The light curve constructed from reported \rxte/PCA and \swift/XRT observations is shown in Figure~\ref{fig:combined_lc} (top).

From the 2007 \swift/XRT non-detection we infer $kT^{\mathrm{\infty}}_{\mathrm{eff}} \lesssim 98$~eV, $F_{\mathrm{X}} \lesssim 5.2\times 10^{-13}~\flux$ and $L_{\mathrm{X}} \lesssim 1.4\times 10^{33}~\dist~\lum$ (0.5--10 keV). On May 31, the X-ray spectrum could be fitted by a power-law model with $\Gamma \simeq 2.5$, suggesting that the source was still accreting at that time \citep[][]{deeg07_hetedecay}. The non-detection could thus not have been more than 7~d into quiescence, and the total off-time no longer than $\simeq$11~d. 

\begin{figure*}
\includegraphics[width=\textwidth]{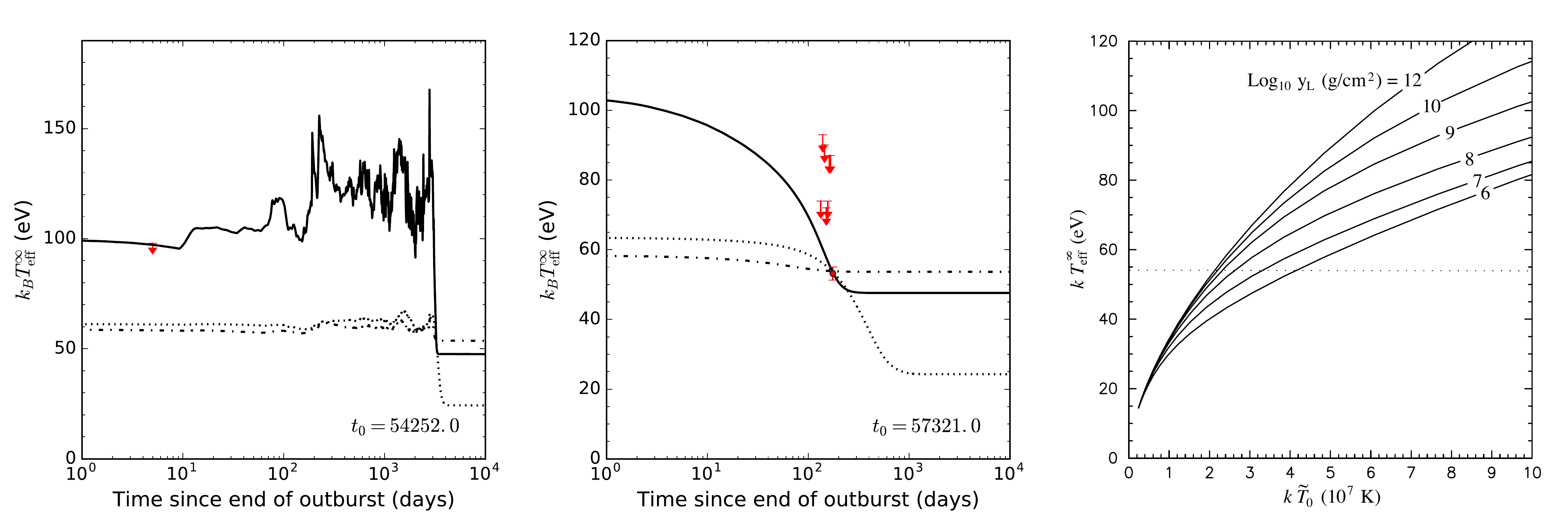}
\caption[]{Left and middle:
Three different thermal evolution simulations of \source\ that account for flux variations during outburst, along with the \chan\ temperature measurement and \swift\ upper limits. On the left the evolution during and after the brief 2007 quiescence epoch and in the middle the evolution in quiescence in 2015--2016. Right: The observed effective temperature $T^\mathrm{\infty}_\mathrm{eff}$
as a function of the red-shifted core temperature $\tilde{T}_0$ (assuming $M=1.4~\Msun$ and $R=10$~km) for varying depth of the light element envelope $y_\mathrm{L}$. The horizontal dotted line shows the 2016 \chan\ measurement of $T^\mathrm{\infty}_\mathrm{eff}$.
}
 \label{fig:models}
 \end{figure*}   

\section{Thermal evolution simulations}\label{sec:model}

Thermal evolution models can be employed to study how the crust of neutron stars is heated during accretion outbursts and subsequently cools in quiescence. 
We use the code \textsc{NSCool} \citep[][]{page2013}, expanded with a new module to incorporate outburst flux variations \citep[][]{ootes2016}. For details we refer to \citet{page2013} and \citet{ootes2016}; here we only discuss the source-specific input. In all our simulations, we require the crust temperature to be consistent with both the 2016 \chan\ and \swift\ data, as well as the 2007 \swift\ upper limit. For consistency with our spectral analysis, we assumed $M=1.4~\Msun$ and $R=10$~km. 

For the outburst properties, we used the publicly available light curves from the \rxte/ASM (2--10~keV), \maxi\ \citep[2--20~keV;][]{maxi2009}, and \swift/BAT \citep[15--50~keV;][]{krimm2013}, which are shown in Figure~\ref{fig:combined_lc} (bottom). The instrument count rates were first converted to Crab units and then to bolometric fluxes ($F^{\mathrm{ob}}_{\mathrm{bol}}$) assuming a correction factor of $c_{\mathrm{bol}}=2$ \citep[][]{galloway06}. The mass-accretion rate was then calculated as $\dot{M}=4 \pi D^2 F^{\mathrm{ob}}_{\mathrm{bol}}/\eta c^2$, where we assumed an accretion efficiency of $\eta=0.2$. This way we estimate an outburst-averaged value of $\langle \dot{M} \rangle \simeq 3.7 \times 10^{-10}~\mdot$ ($\simeq 2.3 \times 10^{16}~\mdotgs$). 
The source was not detected in daily \maxi\ scans after 2015 October 20, implying $L_{\mathrm{X}} \lesssim 5\times10^{35}~\dist~\lum$ (2--20 keV). Based on the rapid flux decay observed in 2007 (Figure~\ref{fig:combined_lc} top), it is plausible that the source entered quiescence quickly after this. We therefore tentatively set the onset of quiescence, $t_0$, to MJD 57321. 

The free parameters in our simulations are the temperature of the neutron star core, which red-shifted value (i.e. in the observer frame) we denote as $\tilde{T}_{0}$, and the thermal conductivity of the crust. The latter is parametrized by the level of impurities $Q_{\mathrm{imp}}$, where a higher $Q_{\mathrm{imp}}$ implies a lower conductivity. Furthermore, we allow for the presence of a source of shallow heat, characterized by a depth $\rho_{\mathrm{sh}}$ and a strength of $Q_{\mathrm{sh}}$, in addition to the energy released in standard nuclear heating reactions. A shallow heat source, with a typical magnitude of $Q_{\mathrm{sh}}\simeq1-2$~MeV and one extreme case of $Q_{\mathrm{sh}} \simeq 6-16$~MeV per accreted nucleon, has been inferred for several neutron stars although its origin remains unknown \citep[e.g.][]{brown08,degenaar2011_terzan5_3,deibel2015,ootes2016,waterhouse2016}. We set $\rho_{\mathrm{sh}} = 4\times10^{8}~\dens$ as determined from modeling the crust cooling curve of \ks\ \citep[][]{ootes2016}, but allowed for different values of $Q_{\mathrm{sh}}$.

We assumed the presence of a thick layer of light elements in the neutron star envelope at a column depth of $y_\mathrm{L} \sim 10^{9}$ g cm$^{-2}$, following the models of \citet{potekhin1997_env}. For a hot star the value of $y_\mathrm{L}$ can have a significant effect on the inferred temperature $\tilde{T}_0$, up to a factor four, and a series of possible set-ups are shown in Figure~\ref{fig:models} (right). For our low temperature of $kT^{\infty}_{\mathrm{eff}} \simeq 54$~eV (dotted line), any $y_\mathrm{L} > 10^{9}$~g~cm$^{-2}$ gives $\tilde{T}_0 \simeq 2\times 10^7$~K.
If $y_\mathrm{L}$ is much smaller than we assume here, $\tilde{T}_0$ could be higher by up to a factor 2.
However, if further cooling occurs, the assumed value of $y_\mathrm{L}$ becomes less important; for $kT^{\infty}_{\mathrm{eff}} \lesssim 30$~eV the inferred $\tilde{T}_0$ is practically independent of $y_\mathrm{L}$ (see Figure~\ref{fig:models} right).

Motivated by crust cooling modeling of other neutron stars \citep[e.g.][]{brown08,page2013,ootes2016}, we started our simulations with $Q_{\mathrm{imp}} = 1$ both with and without shallow heating
(solid and dashed-dotted curves in Figure~\ref{fig:models} left and middle). 
We find that the crust of the neutron star then cools rapidly, within $\simeq$200 days.
Without a shallow heat source (dashed-dotted curve) the crust is hardly heated due to the low outburst accretion rate. 
The current data allow for the presence of a shallow heat source up to $Q_{\mathrm{sh}} \simeq3$~MeV (solid curve). 
We note, however, that for this amount of shallow heating the temperature in 2007 is only just consistent with the \swift\ upper limit (Figure~\ref{fig:models} left). For these two models the 
red-shifted core temperature is $\tilde{T}_{0} = (1.8-2.4)\times10^{7}$~K, corresponding to a final temperature of $kT^{\infty}_{\mathrm{eff}} \simeq 46-53$~eV, i.e. close to our \chan\ measurement. With $Q_{\mathrm{imp}} = 1$, the 
\chan\ point cannot be fitted with a much lower core temperature, hence hardly any further cooling is expected. 

A higher crust impurity keeps the crust hot for a longer period of time. We find that $Q_{\mathrm{imp}} \simeq 8$ is allowed by the data, with no shallow heating needed and a core temperature of $\tilde{T}_{0} = 4\times10^{6}$~K (dotted curve in Figure~\ref{fig:models}). The crust then cools to $kT^{\infty}_{\mathrm{eff}} \simeq 24$~eV in $\simeq$2~yr. 
For $Q_{\mathrm{imp}} > 8$ the temperature curves overshoot our data.

A single impurity parameter for the entire crust is likely not a correct assumption \citep[e.g.][]{page2013}. Theoretically, one expects $Q_{\mathrm{imp}}$ to be high at low densities due to the mixture of elements created during thermonuclear X-ray bursts \citep[e.g.][]{schatz1999,horowitz2007,roggero2016}. However, after crossing the neutron drip density ($\rho_{\mathrm{drip}} \simeq 6\times10^{11}~\dens$), neutrons can be exchanged between nuclei and hence $Q_{\mathrm{imp}}$ likely decreases \citep[e.g.][]{Gupta2008,Steiner2012}. Therefore, we also calculated a model with an impurity parameter that changes at the neutron drip density. We fixed $Q_{\mathrm{imp}} = 1$ at high density and then searched for the maximum allowable impurity parameter at low density that fits the data. The resulting model, without a shallow heat source, has $Q^{\mathrm{low}}_{\mathrm{imp}} = 28$ and $\tilde{T}_{0} = 4 \times10^{6}$~K. The cooling curve looks very similar to our model for $Q_{\mathrm{imp}} = 8$ (dotted curve in Figure~\ref{fig:models}), and therefore we do not show it separately.

Since we cannot fully exclude that \source\ continued to accrete just below the \maxi\ detection limit, we also calculated models with $t_0$ set to MJD 57454, which corresponds to the first \swift/XRT non-detection on 2016 March 7. For $Q_{\mathrm{imp}}=1$, the allowed shallow heating ($Q_{\mathrm{sh}} \simeq 2$~MeV) and core temperature ($\tilde{T}_{0} = 1.5 \times10^{7}$~K) are both lower than for $t_0$ set to MJD 57321, and further cooling to $kT^{\infty}_{\mathrm{eff}} \simeq45$~eV would be expected.

\section{Discussion}\label{sec:discuss}

We report on \chan\ and \swift\ observations obtained within $\simeq$180~d after the $\simeq$10-yr long outburst of the neutron star LMXB and intermittent AMXP \source. Analysis of the \chan\ data reveals that the quiescent spectrum is dominated by soft, thermal emission that can be described by a neutron star atmosphere model with a temperature of $kT^{\mathrm{\infty}}_{\mathrm{eff}} \simeq 54$~eV. Any hard emission tail contributes $\lesssim$9 per cent to the total 0.5--10 keV luminosity of $L_{\mathrm{X}} \simeq 5 \times 10^{31}~\dist~\lum$, for an assumed power-law spectral shape with a photon index of $\Gamma = 1-2$. 

\source\ is different from the AMXPs by exhibiting a thermally-dominated quiescent X-ray spectrum. Other AMXPs (both persistently and intermittently pulsating) display strong power-law emission, contributing $\gtrsim$50\% to the total unabsorbed quiescent 0.5--10 keV flux \citep[e.g.][]{campana2005_amxps,wijnands05_amxps,heinke2009,degenaar2012_amxp}.  

With regard to the quasi-persistent neutron star LMXBs, \source\ is much colder after its long outburst than other sources observed at a similar epoch \citep[$kT^{\mathrm{\infty}}_{\mathrm{eff}} \gtrsim 90$~eV; e.g.][]{wijnands2001,wijnands2002,wijnands2003,wijnands2004,degenaar2010_exo2,fridriksson2010,homan2014}. Moreover, our thermal evolution calculations indicate that the crust of the neutron star in \source\ fully cools in $\simeq$0.5--2~yr, which is short compared to the years-long cooling seen for other quasi-persistent sources \citep[e.g.][]{cackett2013_1659,fridriksson2011,degenaar2014_exo3,homan2014,merritt2016}. 

Our thermal evolution simulations suggest that both the low post-outburst temperature and the relatively fast cooling time scale of \source\ may be due to its relatively low mass-accretion rate during outburst. Indeed, our inferred mass-accretion rate is a factor $\simeq$2 lower than that of the crust-cooling source \exo, and a factor $>$5 lower than that of others \citep[see table~1 in][and references therein]{degenaar2015_ter5x3}. 
For a lower mass-accretion rate, the crust is heated less strongly and its temperature is lower, resulting in a shorter cooling time-scale \citep[see e.g. figure 1 in][]{page2013}. Nevertheless, it is striking that with only a factor $\simeq$2 difference in mass-accretion rate, \exo\ was detected at $kT^{\mathrm{\infty}}_{\mathrm{eff}} \simeq 115$~eV at $\simeq$180 days after its $\simeq$24-yr long outburst \citep[][]{degenaar2010_exo2}. This likely points to other differences between these two sources, e.g. contrasting core temperatures or different crust properties.

Notably, \source\ is the only quasi-persistent LMXB that acted as an AXMP, and the disappearance of its pulsations has been explained as  its magnetic field being ``buried'' by accretion \citep[e.g.][]{cumming2008,patruno2012_hete}. It is currently unclear if/how this should affect the thermal properties. If the magnetic field is strongly folded into the crust
so that it reaches $\simeq 10^{11}$ G, it may create an insulating layer that prevents the accretion-induced heat to propagate to the surface until the magnetic field re-emerges. If so, the neutron star temperature (and perhaps any hard emission) could possibly increase after the Ohmic diffusion timescale. This could be as short as tens of days for \source\ \citep[e.g.][]{cumming2008}. Nevertheless, our work suggests that the quiescent data obtained so far can be explained with similar physical parameters as inferred for other crust-cooling sources. At present there are thus no indications that its magnetic field has a notable impact its thermal properties.

Our simulations show that if the crust of the neutron star is highly conductive with $Q_{\mathrm{imp}} = 1$, no further cooling is expected in \source. Our \chan\ measurement then reflects the temperature of the neutron star core, $\tilde{T}_{0} \simeq 2 \times10^{7}$~K. However, the present data allow for a higher impurity parameter, up to $Q_{\mathrm{imp}} \simeq 8$, for which the core temperature may be as low as $\tilde{T}_{0} \simeq 4\times10^{6}$~K and further cooling of the crust is expected. Future observations of \source\ in quiescence can thus further constrain the impurity impurity of the crust and temperature of the core, as well as possible effects of the magnetic field on the thermal properties.

The total energy released during the $\simeq$10 yr accretion outburst, from both the standard nuclear processes and additional shallow heating, amount to  $\simeq 2\times 10^{43}$~erg. As the present work was in progress, \citet{cumming2016} found that such an amount of energy can raise the core temperature and provides us with a lower limit on its heat capacity that depends on the tantalizing core composition. Comparing with their figure~7, our $\tilde{T}_0 \simeq 2 \times 10^7$ K provides the strongest constraint to date on the core heat capacity: it must be $C>10^{37} (\tilde{T}_{0}/10^8 \, \mathrm{K})$ erg K$^{-1}$, a value that is close to the minimum provided by the leptons in the core. This minimum can only be reached if all baryons are strongly paired, i.e. superfluid or superconducting, hence have a negligible contribution to the total heat capacity. Observing further cooling in \source\ would push this lower limit further down and may limit the maximum fraction of baryons that are paired, and consequently the minimum fraction of baryons that are not paired.

\vspace{-0.2cm}
\section*{Acknowledgements}
We are grateful to the \chan\ team for making this DDT observation possible and to the referee for valuable comments. We acknowledge the use of \swift\ public data archive. ND is supported via an NWO Vidi grant and EU Marie Curie Intra-European fellowship. RW and LO are supported by an NWO Top grant, module 1, awarded to RW. DP is partially supported by the Consejo Nacional de Ciencia y Tecnolog{\'\i}a with a CB-2014-1 grant $\#$240512.

\footnotesize{



}

\end{document}